\documentclass[twocolumn,showpacs,preprintnumbers,amsmath,amssymb]{revtex4}
\usepackage{graphicx}

\newcommand{\figurewidth}{\columnwidth}

\begin{document}

\def\bea{\begin{eqnarray}}
\def\eea{\end{eqnarray}}
\def\a{\alpha}
\def\p{\partial} 
\def\nn{\nonumber}
\def\r{\rho}
\def\xb{\bar{x}}
\def\vb{\bar{v}}
\def\fb{\bar{f}}
\def\lab{\bar{\lambda}}
\def\la{\langle}
\def\ra{\rangle}
\def\f{\frac}
\def\o{\omega}
\def\P{\mathcal{P}}
\def\d{\delta}

\draft

\title{Exact result for nonreciprocity in one-dimensional wave transmission}

\author{Onuttom Narayan$^1$ and Abhishek Dhar$^{1,2}$}
\affiliation{
$^1$Department of Physics, University of California, Santa Cruz, CA 95064\\
$^2$ Raman Research Institute, Bangalore 560080.}
\date{\today}

\begin{abstract}
For sound waves impinging on a one-dimensional medium, we show that
nonlinearity can lead to nonreciprocal transmission, without dissipation
or broken time reversal invariance. Placing quasi-monochromatic filters at
the ends of the nonlinear medium, nonreciprocity can be obtained without
the generation of higher harmonics outside the medium. Remarkably, in
this configuration the nonreciprocity is found to be proportional to
the net energy flow when monochromatic sources of equal strength (at
the filter frequency) are simultaneously turned on at both ends. This
result is conjectured to be general for one dimensional scattering. It is
also shown that although simultaneous monochromatic sources lead to net
energy flow, with sources of small but non-zero bandwidth there is no net
energy transport, in accordance with the second law of thermodynamics.
\end{abstract}


\maketitle
The reciprocity theorem has a long history in acoustics and optics. For
the case of a linear medium with time reversal invariance, the theorem
can be proved~\cite{Rayleigh,Krishnan}: for one dimensional systems, it
amounts to the transmission coefficient being the same when waves are
incident from the left or from the right.  In the absence of time reversal
invariance, reciprocity is no longer necessary. In optics, this can be
achieved by intrinsic magnetization in the scatterer or with an external
magnetic field~\cite{magn}. In acoustics, nonreciprocity can be caused
by --- and be used to detect --- the motion of objects, as in acoustic
tomography~\cite{tomog}.

For nonlinear media, however, even if time reversal invariance
is not broken, nonreciprocity is possible~\cite{lasers}. Photonic
structures with diode like behavior have been proposed, where the
effect of the nonlinearity is strengthened by the existence of a
bandgap~\cite{scalora,konotop}. Such passive diode like behavior can be
useful in the field of optical communications.

In this paper, we examine nonreciprocity in one-dimensional nonlinear
media for longitudinal waves such as sound. 
As with light, we find that nonlinearity is sufficient to result in
nonreciprocity, even for dissipationless systems without broken time
reversal invariance.  
We go on to consider a different configuration: when
monochromatic filters are placed at the two ends of the nonlinear
medium. This confines the higher harmonics to the medium, so that the
reflected and transmitted wave are at the same frequency as the incident
wave. Thus there is no `contamination' from higher harmonics outside
the nonlinear medium. This setup, apart from possible advantages from a
communications perspective, allows us to examine constraints from the
second law of thermodynamics. 

With filters, we obtain the unexpected
result that the nonreciprocity is now proportional to the net energy
transport from one side of the system to the other when two monochromatic
sources of equal strength are simultaneously connected to the two ends of
the device.  We conjecture that this proportionality is {\it general\/}
for any scattering process with two input and two output channels
(at the same frequency) that is invariant under time-translation and
time-reversal and is perturbatively accessible (explained
later in this paper).


We consider a system that can be modelled as two adjacent layers, in each
of which longitudinal waves propagate in accordance with a nonlinear
wave equation.  Outside the system, both to the left and the right,
the linear wave equation is satisfied. Thus we have
\begin{equation}
n_i^2 \ddot y = B_i \partial_x^2 y + \mu_i \partial_x (\partial_x y)^2
\label{quadnl}
\end{equation}
where $y$ is the displacement of the wave, and $n_i,B_i,\mu_i$ vary from
region to region. Inside the scatterer, the two layers have parameters
$(n_1, B_1, \mu_1)$ and $(n_2, B_2, \mu_2).$ Outside the scatterer,
$n = B= 1$ and $\mu = 0.$ The scatterer is taken to cover the region
$-1 < x < 1,$ with the boundary between the two layers at $x=0.$
The form of Eq.(\ref{quadnl}) retains the leading order nonlinearity
in the elasticity of the medium; the energy density of the wave is $
{1\over 2} n^2 \dot y^2 + {1\over 2} (\partial_x y)^2 + \mu
(\partial_x y)^3/3.$ At the three boundaries between the four regions,
$y$ and $B\partial_x y + \mu (\partial_x y)^2$ (the force exerted on 
the boundary from the two regions it separates) are continuous.
We also consider an alternative to Eq.(\ref{quadnl})
\begin{equation}
n_i^2 \ddot y = B_i \partial_x^2 y + \mu_i \partial_x (\partial_x y)^3
\label{tertnl}
\end{equation}
which is slightly easier to work with, but which has an accidental 
$y\rightarrow -y$ symmetry. Eqs.(\ref{quadnl}) and (\ref{tertnl}) are
in the class of Fermi Pasta Ulam (FPU) wave equations~\cite{FPU}.

For both Eqs.(\ref{quadnl}) and (\ref{tertnl}), we first use perturbation
theory to obtain an analytical solution. The incoming wave amplitudes 
from the left and right are $a_1$ and $a_2$ respectively.
Eqs.(\ref{quadnl}) and (\ref{tertnl})
can then be solved 
to linear order in $a_{1,2},$ and then 
iteratively to successive higher orders in perturbation theory. 
For the case without filters, one imposes the requirement that all
frequency components of the solution are purely outgoing outside the
scattering medium, except for the component at frequency $\omega$ whose
incoming part is specified.  For the case with filters, except for the
component at frequency $\omega$ which is unaffected by the filters,
all other components are confined to the scattering medium and have
zero amplitude at $x=\pm 1.$ These conditions are sufficient to solve
Eqs.(\ref{quadnl}) and (\ref{tertnl}), order by order.

The equations were solved to third order using Mathematica$^{\rm TM},$
with $\mu = 1$ and various specific values chosen for $n_{1,2}, B_{1,2}$
and $\omega.$ This third order solution yields the leading $O(|a|^4)$
correction to the outgoing power to the left (or to the right) for
Eq.(\ref{quadnl}).  On the other hand, a similar third order solution
to Eq.(\ref{tertnl}) yields the outgoing power to $O(|a|^6),$ with {\it
two\/} nonlinear contributions to the component at frequency $\omega.$

For the case with filters, the outgoing wave is entirely at frequency
$\omega.$ Expressing the outgoing amplitudes $b_{1,2}$ as an expansion
in powers of $a_{1,2}$ and $a^*_{1,2},$ time translational invariance
requires that each term in the expansion should have one extra power
of the unconjugated variables as compared to the conjugated ones. Thus
$b_i(a_1, a_2) = M_{ij} a_j + N_{ijkl} a^*_j a_k a_l + \ldots.$ The
outgoing power to the left is $|b_1|^2.$ It is possible to verify for
both Eq.(\ref{quadnl}) and Eq.(\ref{tertnl}) that $|b_1(0, a)|^2 -
|b_2(a, 0)|^2$ is not equal to zero, demonstrating nonreciprocity
in the transmission coefficient. (Since the system is nondissipative,
this is equivalent to nonreciprocity in the reflection coefficient.) With
$\mu_{1,2}=1$ and various values of $n_{1,2},$ $B_{1,2},$ and $\omega$ all
$\sim O(1),$ the coefficient of $|a|^4$ in $|b_1(0, a)|^2 - |b_2(a, 0)|^2$
is $O(1).$ In units where $\mu = 1,$ the amplitude of the incoming wave
will be small, so this is a weak effect. However, it should be possible
to enhance the effect by constructing more complicated structures, for
instance acoustic analogs of Ref.~\cite{scalora}.
For the case without filters, one has to consider
the outgoing power at different harmonics separately, but nonreciprocity
is still found.

With filters, one might consider a monochromatic source as a blackbody
(white noise) source from which only waves of one frequency are allowed
to escape\cite{foot2}.  If one connects a blackbody at each end of the
nonlinear scatterer, with both blackbodies at the same temperature,
there should be no net flow of energy from one side to another. With
the filters, this would {\it seem\/} to be equivalent to choosing
$a_{1,2}$ to be equal in magnitude, but with a random relative phase.
Surprisingly, it is possible to verify through the perturbation expansion
of the previous paragraph that if $a_{1,2}$ are indeed of equal strength,
the phase averaged outgoing power is not the same on both sides of the
scatterer. In fact, for {\it all\/} choices of the parameters we have
tried, we have verified that
\begin{eqnarray}
&&|b_1(0, a\sqrt 2)|^2 - |b_2(a \sqrt 2, 0)|^2 \cr
&=& 2 \Big\langle \vert b_1(a, a e^{i\phi})\vert^2\Big\rangle_\phi
- \Big\langle\vert b_2(a, a e^{i\phi})\vert^2 \Big\rangle_\phi.
\label{grand}
\end{eqnarray}
This has been verified to third order for both Eqs.(\ref{quadnl})
and (\ref{tertnl}), i.e. the $O(|a|^4)$ terms for Eq.(\ref{quadnl}),
and the $O(|a|^4)$ and $O(|a|^6)$ terms for Eq.(\ref{tertnl}). We have
been unable to find any result resembling Eq.(\ref{grand}) for the case
without filters, either including or excluding higher harmonics in the
outgoing wave.

Since our analytical treatment is only perturbative, we turn to
numerical simulations.  The nonlinear medium is modelled by a chain of
$N$ particles with anharmonic springs connecting them. Thus if $y_i$
are the displacements of the particles from their equilibrium positions,
\begin{equation}
m_i \ddot y_i = -\partial_{y_i} [V(y_i - y_{i-1}) + V(y_{i+1} - y_i)]
\label{middle}
\end{equation}
for all the particles except the first and the last one, with 
\begin{equation}
V(y) = {1\over 2} y^2 + {\epsilon\over 4} y^4.
\label{potl}
\end{equation}
The first and the last particle must be coupled to the external
environment.  This coupling is through incoming and outgoing waves,
with --- as in any scattering problem --- the incoming waves specified
and the outgoing waves determined by the scattering medium. Beyond the left
boundary of the medium, the external waves can be expressed as $f_i(x -
vt) + f_o(x + vt).$ The force exerted by these waves on the boundary of
the medium is proportional to $-\partial_x[f_i + f_o].$ By continuity, 
the velocity of the boundary is equal to the velocity just outside the 
scattering medium, which is $\partial_t[f_i + f_o].$ From the form
of $f_i$ and $f_o,$ it is 
easy to see that $-v\partial_x[f_i + f_o] = -\partial_t [f_i + f_o] +
2\partial_t f_i.$ Thus the external force acting on the boundary is a
sum of a term proportional to the velocity of the boundary, and a term
specified by the incoming waves.  For monochromatic waves, we have
\begin{eqnarray}
m_1 \ddot y_1    &=& -m_1 \omega_0^2 y_1 
-V^\prime(y_1 - y_2) - \kappa \dot y_1 + A_1\cos (\omega t)\cr
m_N \ddot y_N &=& - m_N \omega_0^2 y_N - V^\prime(y_N - y_{N-1})\cr 
&\qquad& \qquad\qquad-\kappa \dot y_N + A_2\cos(\omega t + \phi)
\label{fl}
\end{eqnarray}
where $\phi$ is the relative phase between the incoming waves from the
left and the right.  Thus the coupling to the external environment is
seen as an effective damping and forcing term in the equation of motion
for the first and last particle.

The first term on the right hand side of Eqs.(\ref{fl}) makes these
particles act as filters if $\omega = \omega_0$ and $m_{1, N}$ are
very large. Due to the nonlinearity of the medium the incoming waves at
frequency $\omega_0$ produce a response at all multiples of $\omega_0.$
If the excitations are resolved into frequency components, for the
component at $\omega_0$ the left hand side of Eqs.(\ref{fl}) cancels the
first term on the right hand side. The forcing and effective damping
term from the external environment must balance the $\partial_y V(y)$
term from the interior, as they would have if the terminal particles
had been missing.  On the other hand, at any higher harmonic, $m_{1,
N}(\omega^2 -\omega_0^2)$ diverges in the $m_{1, N}\rightarrow\infty$
limit, so that $y_{1, N}(n\omega_0)\rightarrow 0$ for $n\neq \pm
1$. Thus for the component at $\omega_0$ the terminal particles are {\it
transparent \/}, whereas for higher harmonics the terminal particles act
as fixed boundaries for $m_{1,N}\rightarrow \infty,$ confining the higher
harmonics to the nonlinear medium.

Eqs.(\ref{middle}) and (\ref{fl}) together with Eq.(\ref{potl}) were
numerically simulated for a chain of $4+2$ particles, in units where
$\kappa = \omega_0 = 1$ and $A_{1, 2} = 1.$ Various values of $m_2\ldots
m_5$ were used; the results shown in Figure 1 are representative. The
nonlinearity parameter $\epsilon$ was varied, which is equivalent to
a fixed $\epsilon$ and varying $A_{1, 2}.$ The chain was started at
$\epsilon = 0,$ allowed to reach steady state, and then $\epsilon$ was
increased slowly till approximately $\epsilon = 0.4,$ after which it was
decreased slowly to zero. The system undergoes a transition as $\epsilon$
is increased, from a periodic state with frequency $\omega = \omega_0 =
1$ to a noisy state. The transition between the two is first order,
with an accompanying hysteresis loop, but for sufficiently small or
large $\epsilon$ only one state is seen.  The existence of two states is
similar to that in Ref.~\cite{diels} for optics.  Perturbation theory,
which is `connected' to 
$\epsilon = 0$ and only allows for harmonics of $\omega_0,$ cannot access
the noisy state. As seen in Figure 1, the perturbatively accessible
periodic state satisfies Eq.(\ref{grand}), whereas the nonperturbative
state does not~\cite{foot5}.
\begin{figure}
\includegraphics[width=\figurewidth]{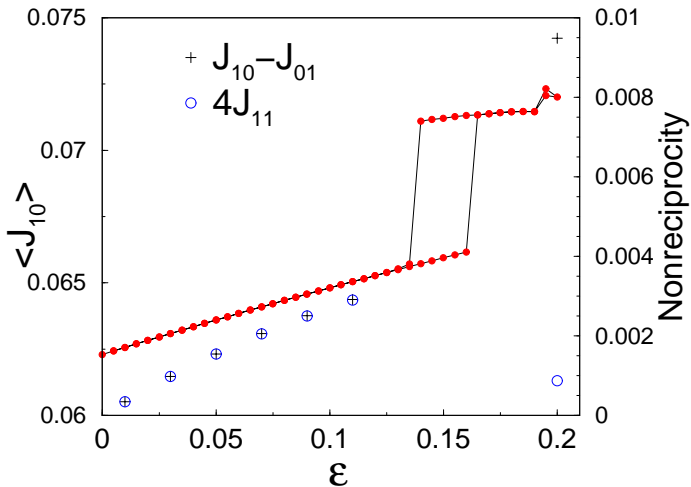}
\caption{ Numerical results for a $4+2$ particle chain with asymmetric
mass-distribution. The particle masses are 100, 1.7, 1.4, 1.9, 1.3
and 100. The left-to-right current $J_{10}=|b_2(1,0)|^2$ is plotted
as a function of the nonlinearity $\epsilon$. The nonreciprocity,
$J_{10}-J_{01}$, is shown for $7$ different values of  $\epsilon$, and
compared with the phase averaged current $J_{11}$  with both sources on
(right vertical scale). The two quantities are equal in the perturbative
state, but not in the non-perturbative state.} 
\label{hyst} 
\end{figure}

In the noisy state, preliminary results when the incident wave is entirely
from the left (on $m_1$) show broad peaks in the transmitted power at
$\omega\approx 0.4\omega_0$ and $\approx 0.15\omega_0.$ As $\epsilon$ is
increased further, there is another transition between $\epsilon\approx
0.35$ and $0.45,$ with a jump in the average transmitted power and a
broadband component to the power spectrum~\cite{foot7}. The jumps from the
perturbative state to the first noisy state and thence to the second noisy
state are at different values of $\epsilon$ when the incident wave comes
from the left instead of the right.  A detailed dynamical analysis would
be required to characterize the various noisy states and the transitions
between them. However, this is not the focus of this paper.

The non-zero right hand side of Eq.(\ref{grand})
might seem to contradict the second law of thermodynamics. If two
blackbody sound sources at the same temperature were connected at the
ends, no net energy flow would be possible. The filters would only allow
waves at frequency $\omega_0$ to enter or exit the system, seemingly
equivalent to monochromatic sources.  However, as can be seen from our
numerical implementation, any filter has a non-zero (albeit arbitrarily
small) bandwidth. For a nonlinear medium, the different frequency
channels interact with each other. Thus even with filters, a blackbody
and monochromatic source are not strictly equivalent~\cite{foot3}. This
may seem like a quibble, but from the discussion before Eqs.(\ref{fl})
it is clear that blackbody sources at the ends would correspond to white
noise being applied to the terminal particles (which is then filtered
by them).  Eqs.(\ref{middle}) and (\ref{fl}) are then generalized
Langevin equations (with damping and noise only in Eq.(\ref{fl})),
which can be rigorously proved to reach thermal equilibrium~\cite{ref1}.
In view of our explicit results for monochromatic radiation, and the
Langevin description for blackbody sources, we must conclude that
narrow and zero bandwidth filters are not equivalent beyond linear
order~\cite{foot1}.  The situation here is different from the one
considered in Ref.~\cite{ONSR}, where non-equilibrium energy sources
were used; since the sources had to be maintained out of equilibrium,
second law arguments were inapplicable there.

We note in passing that Eq.(\ref{middle}) is the standard FPU
system~\cite{FPU}, which is difficult to equilibrate~\cite{Ford}, but
the open boundaries in Eq.(\ref{fl}) seem to be sufficient to equilibrate
the system with thermal (blackbody) sources.

We return to the possible basis of Eq.(\ref{grand}). With filters,
the scatterer can be viewed as generating a mapping from the two
complex input amplitudes to the two complex output amplitudes. This
mapping has to satisfy the properties that i) since the system is
nondissipative, $|b_1|^2 + |b_2|^2 = |a_1|^2 + |a_2|^2$ ii) from time
translation invariance, if $a_{1,2}\rightarrow a_{1,2}e^{i\alpha}$
then $b_{1,2}\rightarrow b_{1,2} e^{i\alpha}$ iii) from time reversal
invariance, if $a_{1,2}\rightarrow b^*_{1,2}$ then $b_{1,2}\rightarrow
a^*_{1,2}$ iv) the mapping is perturbatively accessible from the zero
amplitude limit.  We conjecture that these constraints may be sufficient
to yield Eq.(\ref{grand}). This would imply that the equation is valid
for {\it any\/} one dimensional two-channel scattering problem that
satisfies the conditions above.  In view of our numerical results, the
fourth condition is essential; mappings that violate Eq.(\ref{grand})
can in fact be constructed without it~\cite{mont}.

It is a pleasure to thank Joshua Deutsch, Shyamsunder Erramilli, Richard
Montgomery and Peter Young for several extremely useful discussions. AD
acknowledges support from the NSF under grant DMR 0086287.


\begin{references}
\bibitem{foot2} 
This is not possible without filters, since the incoming wave is at
frequency $\omega,$ while the outgoing wave is at all multiples thereof;
in order to achieve this with a blackbody, one would need a (passive)
one-way filter that would be completely transparent to the higher
harmonics in one direction, {\it irrespective\/} of their intensity,
and opaque in the other direction. Even with the nonreciprocal devices
being considered here, this is not possible.
\bibitem{foot5}
Increasing $\epsilon$ slowly is equivalent to increasing the amplitude of
the incident wave slowly. Alternatively, one can consider the case when
the system starts from rest and the incident wave is switched on. The
transition between the two phases occurs somewhere in the middle of the
hysteresis loop of Figure 1.
\bibitem{foot7}
As $m_{1, N}\rightarrow\infty,$ the noise in the transmitted power
vanishes outside the nonlinear medium. At finite $m_{1, N}$ it is a
suppressed version of the power spectrum inside the medium.
\bibitem{foot1}
For a linear system the different frequencies are independent, 
and there is no non-reciprocity even without time reversal invariance.
A dissipative system may be viewed as one where there are (an infinite
number of) extra channels that the system is coupled to in addition to
those for the waves being scattered, and would not be covered by this 
result.
\end{references}
\end{document}